\documentclass{article}
\usepackage{spconf,amsmath,graphicx}
\usepackage{url}

\title{Multi-objective Optimization Training of PLDA for Speaker Verification}
\name{Liang~He, Xianhong~Chen, Can~Xu and Jia~Liu \thanks{This work was supported by the National Natural Science Foundation of China under Grant No. 61403224, 61370034 and 61273268.}}
\address{Department of Electronic Engineering, Tsinghua University, Beijing, China. \\
 \small{e-mail: \{heliang,chenxianhong,xucan,liuj\}@mail.tsinghua.edu.cn.}}

\begin{document}
\maketitle
\begin{abstract}
Most current state-of-the-art text-independent speaker verification systems take probabilistic linear discriminant analysis (PLDA) as their backend classifiers. The parameters of PLDA are often estimated by maximizing the objective function, which focuses on increasing the value of log-likelihood function, but ignoring the distinction between speakers. In order to better distinguish speakers, we propose a multi-objective optimization training for PLDA. Experiment results show that the proposed method has more than $10\%$ relative performance improvement in both EER and MinDCF on the NIST SRE14 i-vector challenge dataset, and about $20\%$ relative performance improvement in EER on the MCE18 dataset.
\end{abstract}


\section{Introduction}
\label{sec:intro}

Although there are many kinds of frontends, such as GMM/DNN i-vector \cite{Dehak2011,Lei_2014}, TDNN Xvector \cite{Snyder2017Deep}, and DNN embedding \cite{Heigold2015End,Li_rescnn_2017}, probabilistic linear discriminant analysis (PLDA) \cite{Prince2007Probabilistic,Bo2012Face} is still the most popular backend for a text-independent speaker verification system.

Many researchers aim at improving the performance of PLDA \cite{Cumani2013Probabilistic,Burget2011Discriminatively,Rohdin2014Constrained,khosravani2018105}.
Cumani proposes a new PLDA based on i-vector's posterior distribution, where an utterance is not mapped into a single i-vector, but into a posterior distribution to improve the performance for short utterances \cite{Cumani2013Probabilistic}.
Burget refines the PLDA scoring by adopting discriminative models, e.g. support vector machines or logistic regression \cite{Burget2011Discriminatively}.
Following his work, Rohdin introduces more constrains on PLDA parameters to boost system performance \cite{Rohdin2014Constrained}.
Inspired by the nonparametric discriminant analysis \cite{Li2009Nonparametric}, Krosravani proposes a nonparametrically trained PLDA which achieves excellent performance on NIST SRE 2010 core c5 condition \cite{khosravani2018105}.
However, none of the above methods utilizes a discriminant way to \emph{train} space matrix, which plays a crucial role in the PLDA modeling.

We adopt the advantages of discriminant and nonparametric methods, and propose a multi-objective optimization training for PLDA. Experiment results on the NIST SRE14 \cite{sre14} and MCE18 \cite{mce18} demonstrate the effectiveness of proposed methods.

The remainder of this paper is organized as follows.
Section \ref{sec:sgplda} reviews the simplified Gaussian probabilistic linear discriminant analysis (sGPLDA).
Section \ref{sec:motplda} proposes a multi-objective optimization training for sGPLDA.
Section \ref{sec:exp} analyzes and discusses the experiment results.
A conclusion is drawn in Section \ref{sec:con}.

\section{Simplified Gaussian probabilistic linear discriminant analysis} \label{sec:sgplda}
There are many variants of PLDA, the most widely used PLDA in the field of speaker verification is the simplified Gaussian PLDA (sGPLDA) \cite{Bo2012Face} for its simplicity and proven performance on the recent NIST SREs.
The sGPLDA assumes that a length normalized i-vector $\mathbf{x}$ is decomposed into three parts: a global mean vector $\mathbf{\mu}$, a speaker space $F$ and factor $\mathbf{h}_s$, and a $\varepsilon$ which consists of within-class variability and residual noise.
\begin{align} \label{equ:plda}
    \mathbf{x}_{si} = \mathbf{\mu} + F \mathbf{h}_{s} + \varepsilon_{si}
\end{align}
where $s$ is the speaker index and $si$ is the segment index of speaker $s$.
Under the sGPLDA assumption, the speaker factor $\mathbf{h}_{s}$ has a standard normal prior, and $\varepsilon_{si} \sim \mathcal{N}(0, \Sigma_{w})$.
The model parameters $\{F, \Sigma_{w}\}$ and speaker factor $\mathbf{h}_s$ are iteratively optimized by maximizing the log-likelihood function $f$,
\begin{equation}
\begin{aligned} \label{equ:obj}
\mathop{\arg\max}_{F, \Sigma_{w}, \mathbf{h}} f = &  \frac{1}{\sum_{s=1}^S sI} \sum_{s=1}^S \sum_{si=1}^{sI} \\
& \quad \log \mathcal{N}(\mathbf{x}_{si}; \mathbf{\mu} + F\mathbf{h}_{s}, \Sigma_{w}) \mathcal{N}(\mathbf{h}_{s}; 0, I)
\end{aligned}
\end{equation}
via Expectation - Maximization (EM) algorithm \cite{lengthnorm, Bo2012Face}.
$\mathcal{N}$ is a Gaussian distribution, $S$ is the total speaker number and $sI$ is the total segment number of speaker $s$.

\section{multi-objective Optimization Training for sGPLDA}
\label{sec:motplda}
\subsection{Motivation}
Our motivation originates from the linear discriminant analysis (LDA) \cite{Martinez2001lda}.
The LDA is to perform dimensionality reduction by analyzing within-class scatter matrix and between-class scatter matrix. The within-class scatter matrix is
\begin{equation}
\mathbf{S}_w = \sum_{s=1}^S \sum_{si=1}^{sI} (\mathbf{x}_{si} - \mathbf{\mu}_{s}) (\mathbf{x}_{si} - \mathbf{\mu}_{s})^t
\end{equation}
and the between-class scatter matrix is
\begin{equation}
\mathbf{S}_b = \sum_{s=1}^S (\mathbf{\mu}_{s} - \mathbf{\mu}) (\mathbf{\mu}_{s} - \mathbf{\mu})^t
\end{equation}
where $\mathbf{\mu}_{s} = \frac{1}{sI} \sum_{si=1}^{sI} \mathbf{x}_{si}$ is the class mean, and $\mathbf{\mu} = \frac{1}{S}\sum_{s=1}^{S} \mathbf{\mu}_{s}$ is the global mean.

If we analogize sGPLDA and LDA, we will find that
$\mathbf{\mu} + F \mathbf{h}_{s}$ and $\Sigma_{w}$ are equivalent to $\mathbf{\mu}_{s}$ and $\mathbf{S}_w$.
Finding a space $F$ which maximizes $f$ is similar to finding a space $V$ which maximizes $\det{(V^t \mathbf{S}_w V)^{-1}} $.
By these comparison, we find that the objective function $f$ just focuses on within-class vectors, but ignores between-class vectors.
Here, the within-class vectors mean that the vectors are all from the same class and the between-class vectors mean that these vectors are not from the same class. Take an extreme case for example, we only have one speaker's vectors for training. We can compute $F$ by maximizing $f$ and $\mathbf{S}_w$ because they only need within-class vectors, but failed to compute $\mathbf{S}_b$ because it needs between-class vectors.

Clearly, effective use of between-class statistics can further enhance the discriminant ability of designed algorithm, e.g. the space $V$ is obtained by maximizing $\frac{\det{(V^t\mathbf{S}_b V)}}{\det{(V^t\mathbf{S}_w V)}}$ in LDA.
To achieve this goal, we try to integrate between-class statistics into the sGPLDA training.

\subsection{sGPLDA model for between-class vectors}
For a speaker $s$, $\mathbf{x}_{si}$ denotes his/her $i$-th i-vector.
Let $\bar{\mathbf{x}}_{sj}$ denote the $j$-th i-vector that does not belong to speaker $s$, $1 \leq j \leq sJ$.
$\{\mathbf{x}_{si}\}$ and $\{\bar{\mathbf{x}}_{sj}\}$ constitute between-class vectors of speaker $s$, and we use $\mathbf{y}$ to denote them for convenience.
Similar to (\ref{equ:plda}) and (\ref{equ:obj}), $\mathbf{y}_{sk}$ is also decomposed into three parts: a global mean vector $\mathbf{\mu}$, a speaker space $F$ and factor $\mathbf{g}_{s}$, and a $\zeta$ which contains within-class variability of between-class vectors and residual noise.
\begin{align} \label{equ:bv_plda}
    \mathbf{y}_{sk} = \mathbf{\mu} + F \mathbf{g}_{s} + \zeta_{sk}
\end{align}
And the log-likelihood function $g$ is
\begin{equation}
\begin{aligned} \label{equ:bv_obj}
g = & \frac{1}{\sum_{s=1}^S sK} \sum_{s=1}^S \sum_{sk=1}^{sK} \\
& \quad \log \mathcal{N}(\mathbf{y}_{sk}; \mathbf{\mu} + F\mathbf{g}_{s}, \Sigma_{b}) \mathcal{N}(\mathbf{g}_{s}; 0, I)
\end{aligned}
\end{equation}
Here, $sK = sI + sJ$, the factor $\mathbf{g}_{s}$ also has a standard normal prior, and $\zeta_{sk} \sim \mathcal{N}(0, \Sigma_{b})$.
We name (\ref{equ:plda}) and (\ref{equ:obj}) as the sGPLDA model for within-class vectors and (\ref{equ:bv_plda}) and (\ref{equ:bv_obj}) as the sGPLDA model for between-class vectors.

\subsection{Multi-objective Optimization Training}

\begin{figure}[htb]
	\centering
	\centerline{\includegraphics[width=80mm]{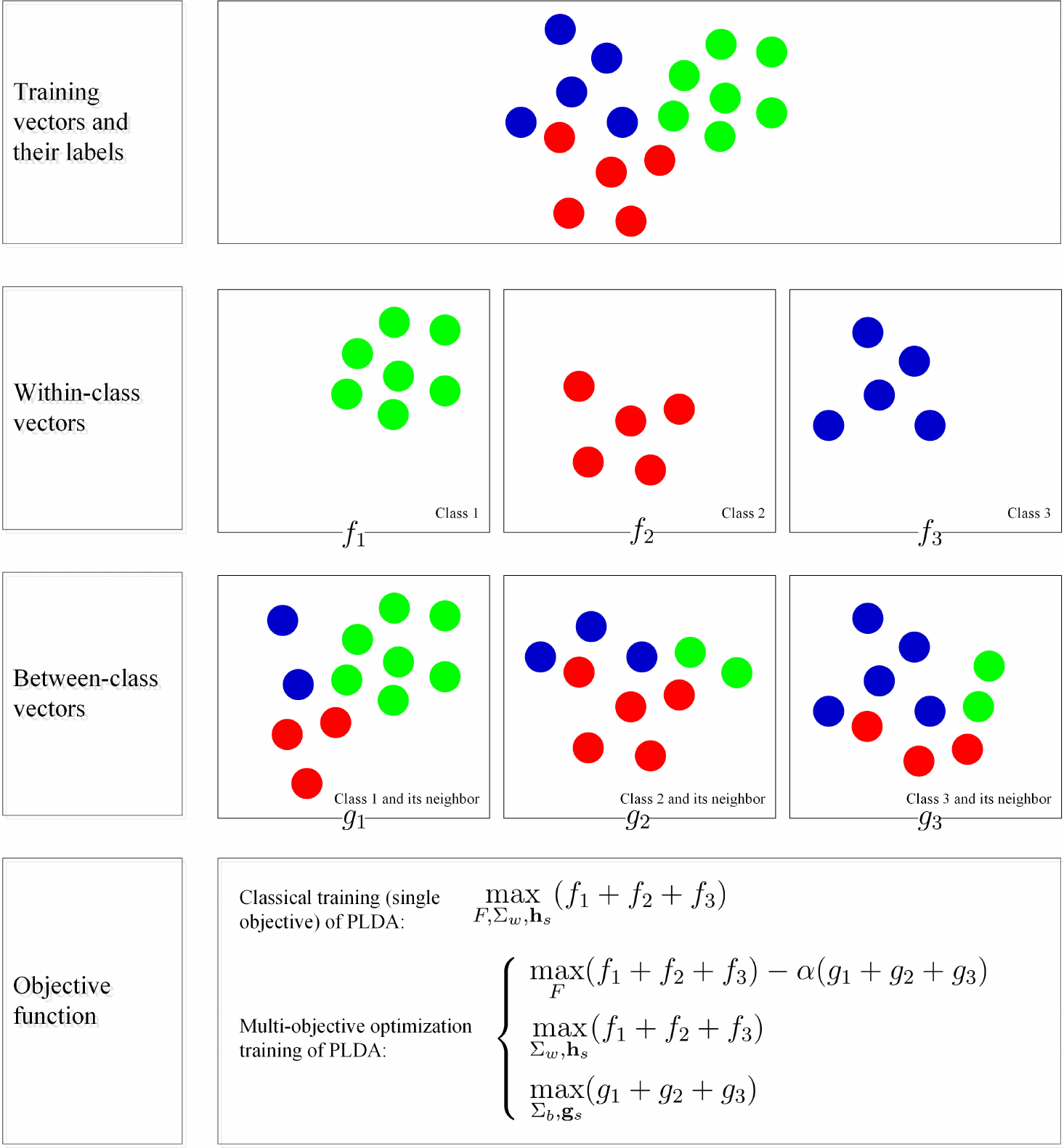}}
	\caption{A demo of multi-objective optimization training of sGPLDA}
	\label{fig:demo}
\end{figure}

The joint model parameters $\{F, \Sigma_{w}, \Sigma_{b}\}$, $\mathbf{h}_s$, and $\mathbf{g}_s$ are obtained by multi-objective optimization training, see Fig. \ref{fig:demo}.
Our considerations are as follows:

\begin{enumerate}
    \item The sGPLDA model for within-class and between-class vectors share the same speaker space $F$. Intuitively, the desired $F$ is to maximize $f$ and to minimize $g$ at the same time. Therefore, the objective function is $\mathop{\arg\max}_{F} (\alpha f - g)$, where $\alpha$ is an introduced factor which balances $f$ and $g$, and will be examined in the experiment section.
    \item $\Sigma_{w}$ and $\mathbf{h}_s$ are only related to $f$, and the objective function is $\mathop{\arg\max}_{\Sigma_{w}, \mathbf{h}_s} (f)$.
    \item $\Sigma_{b}$ and $\mathbf{g}_s$ are only related to $g$, and the objective function is $\mathop{\arg\max}_{\Sigma_{b}, \mathbf{g}_s} (g)$.
%    \item $\mathbf{\mu}$ is the global mean vector which is given by $\mathbf{\mu} = \frac{\sum_{s=1}^S \sum_{si=1}^{sI} \mathbf{x}_{si}}{\sum_{s=1}^S \sum_{si=1}^{sI} 1}$.
\end{enumerate}

The parameters are obtained by EM algorithm.
During the E-step, $\mathbf{h}_s$ and $\mathbf{g}_s$ are obtained by taking the derivation with $f$ and $g$, respectively.
\begin{equation}
    \begin{aligned}
        \mathbf{h}_s & = [(sI) F^t \Sigma_{w}^{-1} F + I]^{-1} F^t \Sigma_{w}^{-1} \sum_{si=1}^{sI} (\mathbf{x}_{si} - \mu) \\
        \mathbf{g}_s & = [(sK) F^t \Sigma_{b}^{-1} F + I]^{-1} F^t \Sigma_{b}^{-1} \sum_{sk=1}^{sK} (\mathbf{y}_{sk} - \mu)
    \end{aligned}
\end{equation}
During the M-step, $F$$, \Sigma_{w}$, and $\Sigma_{b}$ are obtained by taking the derivation with $\alpha f - g$, $f$, and $g$, respectively.
\begin{equation}
    \begin{aligned}
        F = & \bigg(\frac{\alpha}{\sum_{s=1}^{S} sI} \sum_{s=1}^S \sum_{si=1}^{sI} (\mathbf{x}_{si} - \mathbf{\mu}) \mathbf{h_s}^t \\
        & - \frac{1}{\sum_{s=1}^{S} sK} \sum_{s=1}^S \sum_{sk=1}^{sK} (\mathbf{y}_{sk} - \mathbf{\mu}) \mathbf{g_s}^t \bigg)  \\
        & \bigg( \frac{\alpha}{\sum_{s=1}^{S} sI}\sum_{s=1}^S (sI) (\mathbf{h}_s \mathbf{h}_s^t) \\
        & - \frac{1}{\sum_{s=1}^{S} sK}\sum_{s=1}^S (sK) (\mathbf{g}_s \mathbf{g}_s^t) \bigg)^{-1}
    \end{aligned}
\end{equation}
and
\begin{equation}
    \begin{aligned}
        \Sigma_{w} &=\frac{1}{\sum_{s=1}^{S} sI} \sum_{s=1}^S \sum_{si=1}^{sI} (\mathbf{x}_{si}-\mathbf{\mu}-F \mathbf{h}_s) (\mathbf{x}_{si}-\mathbf{\mu}-F \mathbf{h}_s)^t \\
        \Sigma_{b} &=\frac{1}{\sum_{s=1}^{S} sK} \sum_{s=1}^S \sum_{sk=1}^{sK} (\mathbf{y}_{sk}-\mathbf{\mu}-F \mathbf{g}_s) (\mathbf{y}_{sk}-\mathbf{\mu}-F \mathbf{g}_s)^t
    \end{aligned}
\end{equation}
The E-step and M-step are iteratively performed.

\subsection{Selection of $\bar{\mathbf{x}}_{sj}$}
As stated in \cite{He2018Local}, the most challenging task in text-independent speaker verification is to discriminate easily confusable speakers. Krosravani also proposes a nonparametrically trained PLDA, in which the core idea is selecting nearest neighbor vectors during scoring \cite{khosravani2018105}.
Therefore, we adopt random and nearest neighbor selections to pick up $\bar{\mathbf{x}}_{sj}$.
The former is used for comparison and we believe the latter is effective.
The nearest neighbor selection is that for a speaker $s$, we calculate inner products between $\mathbf{x}_s$ and $\bar{\mathbf{x}}_{sj}$, sort them in a descending order and select the top $sI$ $\bar{\mathbf{x}}_{sj}$.

\subsection{Verification score}
The scoring is also based on two-covariance model \cite{Bo2012Face} and the log-likelihood ratio is
\begin{equation}
    \begin{aligned}
        score(\mathbf{x}_1, \mathbf{x}_2) = &\log \frac {p(\mathbf{x}_1,\mathbf{x}_2|\text{same speaker})} {p(\mathbf{x}_1,\mathbf{x}_2|\text{different speakers})} \\
        =& \mathbf{x}_{1}^t Q \mathbf{x}_{1} + \mathbf{x}_{2}^t Q \mathbf{x}_{2} +
        2 \mathbf{x}_{1}^t P \mathbf{x}_{2} + \text{const}
    \end{aligned}
\end{equation}
where
\begin{equation}
    \begin{aligned}
        Q &= \Sigma_{tot,b}^{-1} -
        (\Sigma_{tot,w} - \Sigma_{ac}\Sigma_{tot,w}^{-1}\Sigma_{ac})^{-1} \\
        P &= \Sigma_{tot,w}^{-1} \Sigma_{ac} (\Sigma_{tot,w} - \Sigma_{ac}\Sigma_{tot,w}^{-1}\Sigma_{ac})^{-1} \\
        \Sigma_{tot,w} &= FF^t + \Sigma_{w}, \Sigma_{tot,b} = FF^t + \Sigma_{b}, \text{and } \Sigma_{ac} = FF^t
    \end{aligned}
\end{equation}
Different from \cite{Bo2012Face}, we use $\Sigma_{tot,b}$ instead of $\Sigma_{tot,w}$ to compute $Q$, because $\Sigma_{tot,b}$ is a more reasonable choice under the different speakers assumption.

\section{Experiments}
\label{sec:exp}
\subsection{NIST i-vector Machine Learning Challenge, SRE14}
NIST i-vector machine learning challenge (SRE14) takes i-vectors instead of speech as input to examine the backend of speaker verification system \cite{sre14}.
It is gender independent, contains 1306 speaker models, 9634 test segments and 12582004 trials.
Each speaker model has 5 i-vectors.
The trials are randomly divided into a progress subset ($40\%$) and an evaluation subset ($60\%$).
In addition, NIST provided a development set, containing 36572 i-vectors.
All the i-vectors are 600-dimensional.
We study the backend learning algorithms with development labels known.
After applying LDA, traditional single objective function (SO) sGPLDA and multi-objective optimization training of sGPDLA (MO) are comparatively studied under the same condition.
The dimension of LDA, SO sGPLDA, and MO sGPLDA are 250, 150, and 150, respectively.
Unless otherwise specified, $\bar{\mathbf{x}}_{sj}$ is nearest selected.

\begin{figure}[htb]
    \centering
    \centerline{\includegraphics[width=90mm]{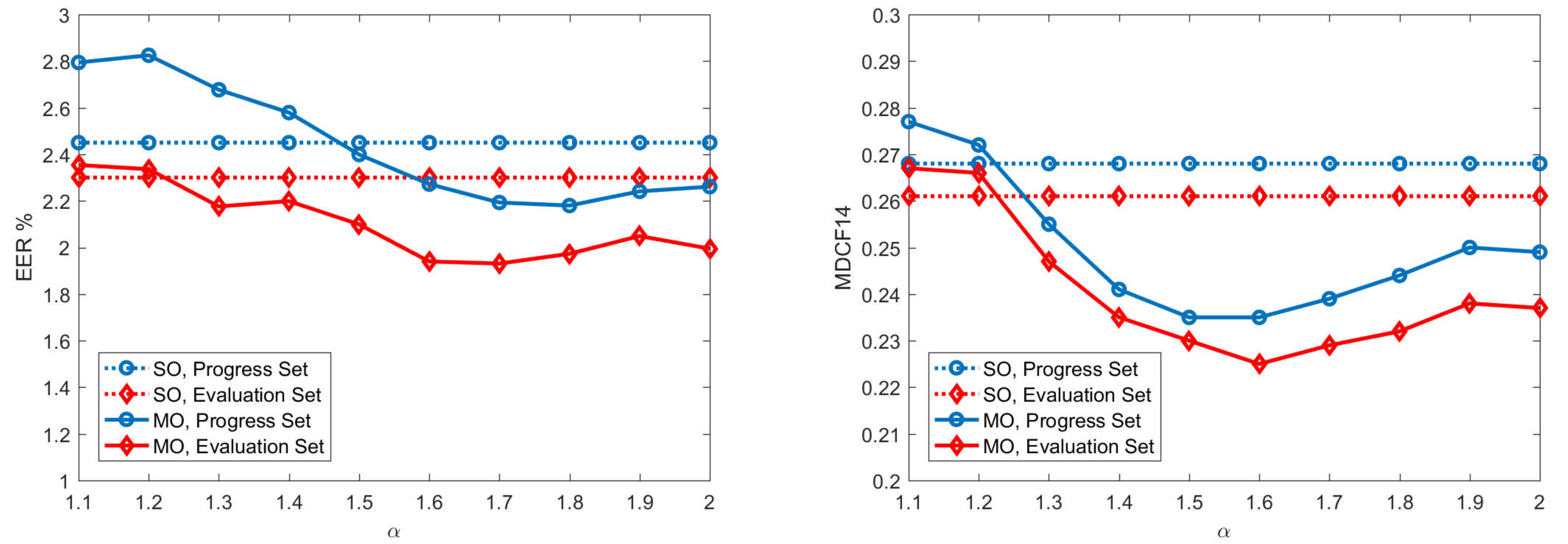}}
    \caption{The EER and MDCF14 vary with the $\alpha$}
    \label{fig:alpha}
\end{figure}

\begin{figure}[htb]
    \centering
    \centerline{\includegraphics[width=90mm]{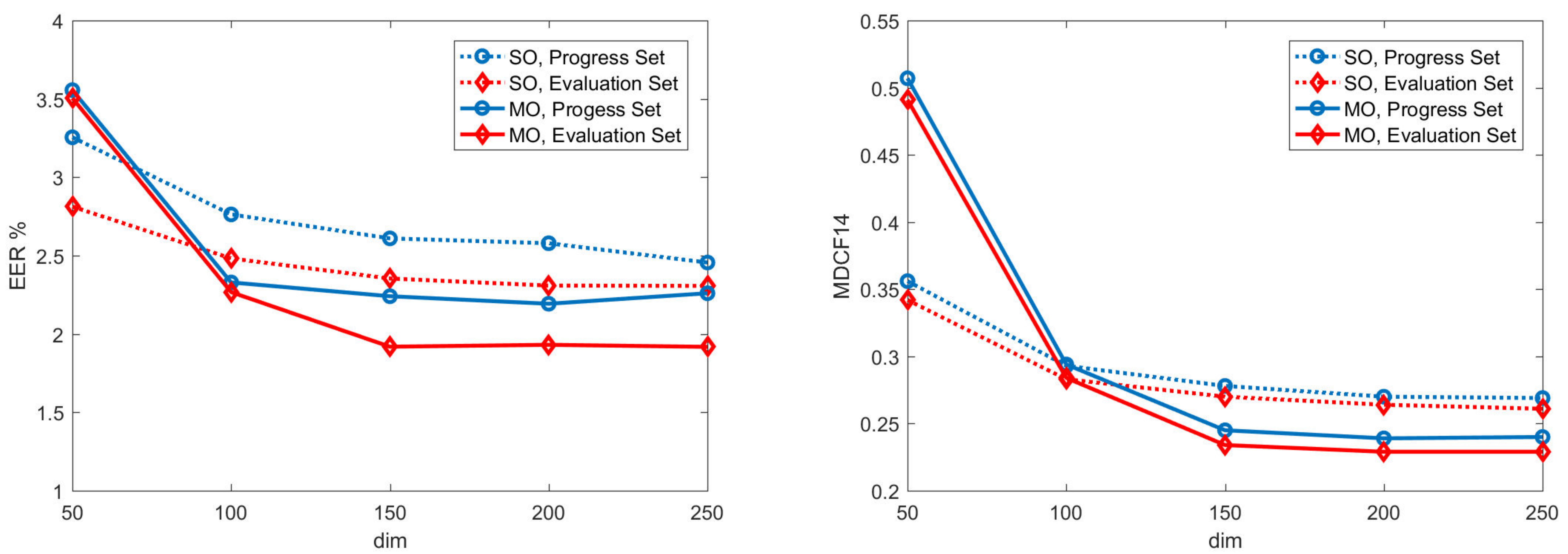}}
    \caption{The EER and MDCF14 vary with the dimension}
    \label{fig:dim}
\end{figure}

Fig.\ref{fig:alpha} shows that the EER and MDCF14 of SO/MO sGPLDA vary with the $\alpha$.
It can be seen that as $\alpha$ changes from $1.1$ to $2$ with a step of $0.1$, both EER and MDCF14 decrease first and then increase, which means a well balance between $f$ and $g$ is important for MO sGPLDA.
We choose $\alpha = 1.7$ in the following experiments.

Fig.\ref{fig:dim} shows that the EER and MDCF14 of SO/MO sGPLDA vary with the sGPLDA dimension.
In most cases ($150$, $200$, and $250$), MO sGPLDA outperforms SO sGPLDA.
At a low dimension ($50$), the performance of MO sGPLDA is decreased.

\begin{figure}[h]
	\centering
	\centerline{\includegraphics[width=85mm]{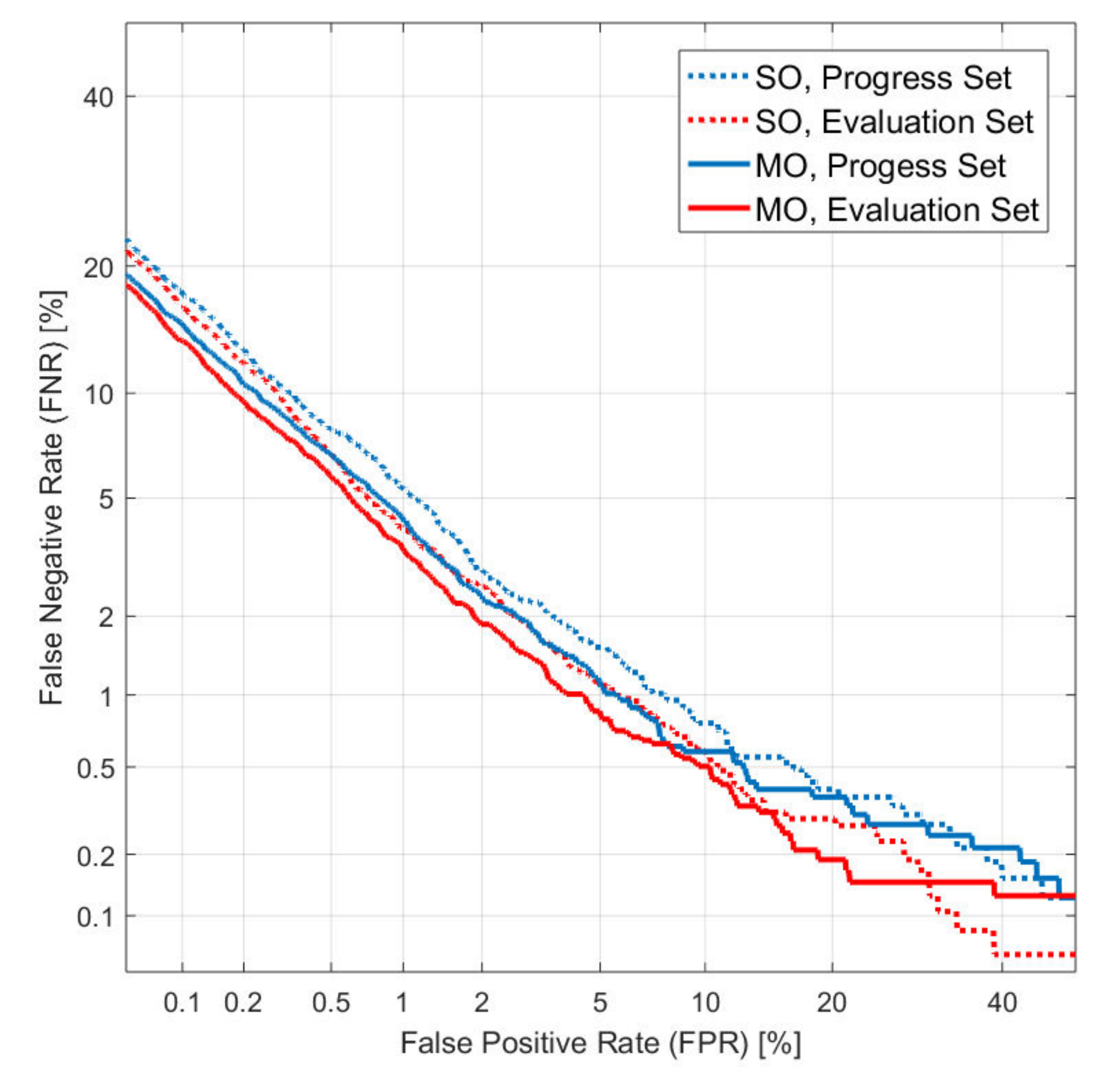}}
	\caption{The DET curves of sGPLDA on NIST SRE14}
	\label{fig:det}
\end{figure}

From Table \ref{tbl:results}, we can see that nearest neighbor selection is better than random selection, which is in line with our expectation.
The nearest selected $\bar{\mathbf{x}}_{sj}$ are easily confusable i-vectors.
Taking them into considerations can boost system performance.

Compared with the SO sGPLDA, both EER and MDCF of proposed MO sGPLDA with nearest selection have more than $10\%$ relative improvement on the progress set and the evaluation set of NIST SRE14, see Table \ref{tbl:results} and Fig.\ref{fig:det}.

\begin{table}[!h]
	\begin{center}
		\caption{Experiment results of sGPLDA on NIST SRE14.}
		\label{tbl:results}
		\begin{tabular}{|c|c|c|}
			\hline
			& EER[\%]    & MDCF14 \\
			\cline{2-3}
			& \multicolumn {2}{|c|}{Progress Set}       \\
			\hline
			SO       & 2.45	         & 0.268      \\
			\hline
			MO,Random    & 3.59	 & 0.319   \\
			\hline
			MO,Nearest    & \bf{2.19}	 & \bf{0.239} \\
			\hline
			& \multicolumn {2}{|c|}{Evaluation Set}       \\
			\hline
			SO    & 2.30	    & 0.261     \\
			\hline
			MO,Random    & 3.04	& 0.304   \\
			\hline
			MO,Nearest   & \bf{1.93}	& \bf{0.229} \\
			\hline
		\end{tabular}
	\end{center}
\end{table}

\subsection{MCE18}

The 1st Multi-target speaker detection and identification Challenge Evaluation \cite{mce18} provides three i-vector sets: training, development and test sets. Each set consists of blacklist and non-blacklist (background) speakers.

For the training set, there are 3,631 blacklist speakers and 5,000 background speakers.
Each blacklist speaker has 3 i-vectors, and there are 10,893 i-vectors for blacklist speakers in total.
For the development set, there also 3,631 blacklist speakers and 5,000 background speakers.
Each speaker has only one i-vector.
The blacklist speakers of the training and development sets are the same while the background speakers are not.
No information is provided about the distribution of speakers in the test set.
All the i-vectors are 600 dimension.
The MCE18 evaluation dataset includes the Fixed and Open conditions. In the Fixed condition, we can only use data provided by the MCE18.
This limitation is removed in the Open condition.
We examined the Mot sGPLDA on the Fixed condition test.
Our procedure is classical, includes length normalization \cite{lengthnorm}, LDA, PLDA and score normalization in turn.
We use both training and development sets to train these parameters.
The dimension of both LDA and PLDA is 350.

From the Table \ref{tbl:mce14}, we can see that, compared with the SO sGPLDA, the proposed MO sGPLDA has $19.8\%$ and $22.0\%$ relative improvement in the \emph{Top S} and \emph{Top 1} conditions on the MCE18 evaluation dataset, which further proves our assert that the parameters trained by multi-objective optimization training not only have the ability to maximize the log-likelihood function on the within vector sets, but also have the ability to distinguish the vectors which are easily mis-judged.

\begin{table}[!h]
	\begin{center}
		\caption{Experiment results of sGPLDA on MCE18.}
		\label{tbl:mce14}
		\begin{tabular}{|c|c|c|}
			\hline
			EER[\%]       & Top S    & Top 1 \\
			\hline
			SO            & 6.75	 & 9.39  \\
			\hline
			MO,Nearest    & \bf{5.41} & \bf{7.32} \\
			\hline
		\end{tabular}
	\end{center}
\end{table}

%\vfill\pagebreak
\section{Conclusion}
\label{sec:con}
We propose a multi-objective optimization training for the sGPLDA.
It not only focuses on increasing the log-likelihood function, but also improves the distinction ability between easily mis-judged speakers.
Compared with the traditional method, the EER and MDCF of multi-objective optimized sGPLDA have $10.5\%$ and $11.1\%$ relative performance improvements on SRE14 progress set, and $16.2\%$ and $12.1\%$ relative performance improvements on SRE14 evaluation set, and the EER of multi-objective optimized sGPLDA have $19.8\%$ and $22.0\%$ relative performance improvements in the \emph{Top S} and \emph{Top 1} conditions on the MCE18 evaluation set.

This method can also be extended to other types of PLDA with proper modification. The python and matlab code for this paper can be downloaded from Github: \textit{git clone https://github.com/sanphiee/MOT-sGPLDA-SRE14} and \textit{git clone https://github.com/sanphiee/MOT-sGPLDA-MCE18}.

\vfill\pagebreak

\bibliographystyle{IEEEbib}
\bibliography{reference_for_motplda}

\end{document}